\documentstyle{article}
\def\makefront{\vspace*{1cm}\begin{center}
\def\newtitleline{\\ \vskip 5pt}
{\Large\bf\titleline}\\
\vskip 1truecm
{\large\bf\authors}\\
\vskip 5truemm
\addresses
\end{center}
\vskip 1truecm
{\bf Abstract:}
\abstracttext
\vskip 1truecm}
\def\be{\begin{equation}}
\def\ee{\end{equation}}
\def\bc{\begin{center}}
\def\ec{\end{center}}
\def\bea{\begin{eqnarray}}
\def\eea{\end{eqnarray}}

\def\dmsq{{(\Delta m^2)}}

\def\ev{{\rm \; eV}}
\def\gev{{\rm \; GeV}}

\def\grav{\tilde{G}}

\def\ls{{\Lambda_S}}

\def\mt{{\tilde{m}_f}}
\def\mtsq{{\tilde{m}_f^2}}
\def\ov{\overline}

\def\psie{f}
\def\psieb{\overline{f}}
\def\psis{\tilde{G}}
\def\psisb{\overline{\tilde{G}}}

\def\te{\tilde{f}}
\def\tg{\tilde{G}}

\begin {document}
\vspace*{-2.0cm}
hep-th/9801003 \hfill{DFPD~98/TH/02}
\def\titleline{
Effective interactions of a light gravitino
\footnotemark[1]}
\footnotetext[1]{Talk given at the Workshop {\em Quantum 
Aspects of Gauge Theories, Supersymmetry and Unification} 
(Neuch\^atel University, Neuch\^atel, Switzerland, 18-23 
September 1997), to appear in the Proceedings} 
\def\authors{Fabio Zwirner}
\def\addresses{
Istituto Nazionale di Fisica Nucleare, Sezione di Padova,
\\
Dipartimento di Fisica, Universit\`a di Padova, 
\\
Via Marzolo~8, I-35131 Padua
}
\def\abstracttext{
We review some recent results on the effective interactions of
a light gravitino with ordinary particles. In particular, we
discuss on a simple example a novel aspect of the low-energy 
theorems for broken supersymmetry: in the effective lagrangian 
describing the goldstino couplings to matter, there are terms 
bilinear in the goldstino that, already at the lowest non-trivial 
order, are not entirely controlled by the supersymmetry-breaking 
scale, and introduce additional free parameters. We conclude by 
mentioning some phenomenological implications, including a 
lower bound on the gravitino mass from collider data.
}
\makefront
The interest in the low-energy effective interactions of the
gravitino (goldstino) in the spontaneously broken phase dates 
back to the early days of supersymmetry \cite{va,dwf,fayet}. 
Since then, significant theoretical progress was made \cite{nl}. 
However, some aspects of the subject have been clarified only 
recently \cite{bfz3,bfzeegg}, following a renewed interest (see, 
e.g., \cite{phenorec} and references therein) in the physics at 
energies much larger than the gravitino mass, but smaller than 
the masses of the other supersymmetric particles. This talk 
begins with some introductory remarks on spontaneously broken 
$N=1$ supersymmetry and its low-energy limit, continues with 
the illustration of some recent theoretical developments 
\cite{bfz3,bfzeegg}, and ends by mentioning some of 
their phenomenological implications, whose exploration
has just started \cite{bfz4,bfmz}.
\section{The general framework}

Most of the participants in this Workshop share the view
that the fundamental theory lying beyond the Standard Model 
should exhibit a spontaneously broken $N=1$ supersymmetry. 
Many dynamical questions on spontaneous supersymmetry 
breaking are still unanswered, but the `kinematical' aspects 
of the problem are well understood \cite{global,local}. 
We begin our discussion by recalling the supergravity potential:
\be
V = ||F||^2 + ||D||^2 - ||H||^2 \, ,
\ee
where $||F||^2$, $||D||^2$ and $||H||^2$ are positive-definite
quantities, controlled by the auxiliary fields of the chiral,
vector and gravitational supermultiplets, respectively. The
microscopic scale of supersymmetry breaking and the gravitino 
mass are given by $\ls \equiv \langle ||F||^2 + ||D||^2  
\rangle^{1/4}$ and $m_{3/2} \equiv \langle ||H|| \rangle / 
(\sqrt{3} M_P)$, respectively, where $M_P \equiv (8 \pi 
G_N)^{-1/2} \simeq 2.4 \times 10^{18} \gev$ is the Planck 
mass. From the facts that we live in an approximately flat 
space-time and that no direct signal of supersymmetry has 
been detected so far, 
\be
\Lambda_S^2 \simeq \sqrt{3} \, m_{3/2} \,  M_P \, ,
\ee
with an extremely good accuracy. The VEVs of the auxiliary 
fields of the chiral and vector multiplets identify the 
goldstino, which provides the $\pm 1/2$ helicity components 
of the massive gravitino:
\be
\grav = \langle F_i \rangle \psi^i + 
\langle D_a \rangle \lambda^a \, .
\ee
The generic form of the supersymmetry-breaking mass splittings
is
\be
\dmsq_I \sim g_I \cdot \Lambda_S^2 \simeq g_I 
\cdot \left( \sqrt{3} \, m_{3/2} \, M_P \right) \, ,
\ee
where $g_I$ is the effective coupling of the goldstino
multiplet to the sector `I' of the spectrum. Even if we 
stick to the prejudice that these splittings are of the 
order of the electroweak scale, for supersymmetry to 
play a r\^ole in the solution of the hierarchy problem, 
this does not fix $\ls$ (or, equivalently, $m_{3/2}$), 
since the couplings $g_I$ are model-dependent. The dynamical 
origin of the scales $\ls$ and $m_{3/2}$ is still obscure, 
and different possibilities can be legitimately considered.
In this talk we concentrate on the case where the 
interactions of the goldstino multiplet are much stronger 
than the gravitational ones, so that the gravitino is 
much lighter than all the other supersymmetric particles. 
Our goal is to discuss the low-energy limit, with the typical
energies of the processes under consideration much larger 
than the gravitino mass, but smaller than the masses of all 
the other particles not belonging to the Standard Model.

A rather explicit method that can be used for this study is 
based on two simple logical steps. First, we start from a
generic supergravity lagrangian, assuming that supersymmetry 
is spontaneously broken with a light gravitino, and we take 
the appropriate low-energy limit: $M_P \to \infty$ with $\ls$ 
fixed. In accordance with the supersymmetric equivalence 
theorem \cite{fayet,susyeq}, gravitational interactions are 
consistently neglected in this limit, and we end up with an 
effective (non-renormalizable) theory with linearly realized, 
although spontaneously broken, global supersymmetry, whose 
building blocks are the chiral and vector supermultiplets
containing the light degrees of freedom, including the goldstino. 
Since no supersymmetric partners of ordinary particles have been
observed yet, a situation of present interest is the one where the
available energy is smaller than the supersymmetry-breaking
mass splittings. In such a context, we can perform the second
step and move to a `more effective' theory, by explicitly 
integrating out the heavy superpartners in the low-energy limit. 
The only degrees of freedom left are then the goldstino and the 
Standard Model particles, and supersymmetry is non-linearly 
realized. 

We now discuss what can be learnt by comparing the results 
obtained via the previous method with the conventional 
wisdom \cite{va,nl} on non-linear realizations: on-shell 
scattering amplitudes with two goldstinos and additional 
matter or gauge particles are fully controlled, at the 
leading non-trivial order in the low-energy expansion, by 
the scale $\ls$ of supersymmetry breaking (with no further 
reference to the supersymmetry-breaking mass splittings), and 
by the canonical energy-momentum tensor $T_{\mu \nu}$ of the 
matter and gauge fields. We list here a number of interesting 
cases where this comparison can be done.

\begin{itemize}
\item
The amplitudes involving two goldstinos and two photons:
$\grav \grav \gamma \gamma$. It was believed for some time
that the processes described by these amplitudes could be 
relevant for stellar cooling, and may allow to put an interesting 
lower bound on $m_{3/2}$.
After some controversial results, however, it was found 
\cite{bfz3} that the amplitudes obtained via the explicit 
construction outlined above are equivalent to those obtained 
from the conventional effective lagrangian of the non-linear 
realization. The energy suppression is then so strong that 
only very weak bounds on $m_{3/2}$ can be derived. 
\item
The amplitudes involving two goldstinos and two
matter fermions: $\grav \grav f \ov{f}$. These 
are the ones that have recently revealed \cite{bfzeegg}, 
via some puzzling results, some previously unnoticed
features of the low-energy theorems for supersymmetry:
they will be discussed in detail, on a simple example, 
in the rest of this talk. Also the processes described 
by these amplitudes were believed to play a r\^ole in 
stellar cooling but, as we shall see, no interesting
bound on $m_{3/2}$ can be obtained either \cite{bfzeegg}.
\item
The amplitudes involving two goldstinos, two matter 
fermions and a vector boson, for example the photon 
or the gluon: $\grav \grav f \ov{f} \gamma$, $\grav 
\grav f \ov{f} g$. They are the most important ones 
for phenomenology. In the case of heavy superpartners, 
they can lead to an absolute lower bound on $m_{3/2}$
or, optimistically, to the best signals for the discovery 
of supersymmetry. At the end of this talk, we shall mention
some recent results on the phenomenology of light gravitinos 
at $e^+ e^-$ \cite{bfz4} and hadron \cite{bfmz} colliders.
\end{itemize}

\section{A puzzling result}

Consider, for simplicity, a globally supersymmetric theory 
with only two chiral superfields, one describing the 
goldstino $\tg$ and its complex spin-0 partner $z \equiv (S + 
i P) / \sqrt{2}$, the other one describing a massless left-handed 
matter fermion $f$ and its complex spin-0 partner $\te$. According 
to the standard formalism \cite{wb}, and neglecting for the moment 
higher-derivative terms, the lagrangian is completely specified in 
terms of a superpotential $w$ and a K\"ahler potential $K$. Assume 
that, at the minimum of the potential, $\langle z \rangle \! = 0$, 
$\langle \te \rangle \! = \langle  F^{1}  \rangle = 0$ (consistently 
with matter conservation), and $\langle  F^0  \rangle \ne 0$, where 
$F^0$ and $F^1$ denote the auxiliary fields associated with $\grav$
and with $f$, respectively. 

Our goal is to identify the effective four-fermion interaction
involving two matter fermions and two goldstinos. Expanding around 
the vacuum, we can write:
\be
w=\hat{w}(z)+ \ldots \, , 
\;\;\;\;\;
K=\hat{K}(z,\bar{z})+\tilde{K}(z,\bar{z}) \, |\te|^2
+\ldots \, ,
\label{expan}
\ee
where, here and in the following, the dots denote terms that are 
not relevant for our considerations. The spectrum can be easily 
derived from standard formulae \cite{wb}. The goldstino and the 
matter fermion remain massless, whilst all the spin-0 particles 
acquire in general non-vanishing masses. Moreover, the expansion 
of the lagrangian in (canonically normalized) component fields can 
be rearranged in such a way that all the interaction terms relevant 
for our calculation are expressed in terms of the mass parameters 
$(m_S^2,m_P^2,\mtsq)$, associated with the spin-0 partners of the 
goldstino and of the matter fermion, and the scale $\ls$ of 
supersymmetry breaking, without explicit reference to $w$ and $K$:
\be
{\cal L} = 
- 
{1\over 2\sqrt{2} F} [ (m_S^2 S + i m_P^2 P) \psis 
\psis + {\rm h.c.} ]
- {\mtsq \over F} (\te^* \, \psis \psie +\te \, \psisb \psieb) 
-{\mtsq \over F^2} \, \psis \psie \, \psisb \psieb + \ldots \, .
\label{lag}
\ee
In (\ref{lag}), we have used two-component spinors. For simplicity,
$F \equiv \, < \! \ov{w}_{\ov{z}} (K_{\ov{z} z})^{-1/2} \! >$ 
(lower indices denote derivatives), which defines here the 
supersymmetry-breaking scale $\ls = |F|^{1/2}$, was assumed to be real. 

Starting from (\ref{lag}), we take the limit of a heavy 
spin-0 spectrum, with $(m_S,m_P,\mt)$ much larger than the 
typical energy of the scattering processes we would like to study. 
In this case, we can build an effective lagrangian for the light 
fields by integrating out the heavy states. As discussed in detail
in \cite{bfz3}, the crucial property of such an effective lagrangian 
is its dependence on the supersymmetry-breaking scale $F$ only, 
without any further reference to the supersymmetry-breaking masses 
$(m_S,m_P,\mt)$. This property is the result of subtle cancellations 
among different diagrams, corresponding to the contact interaction
associated with the last term in eq.~(\ref{lag}) and to $\te$ 
exchange, and agrees with previous results \cite{va,dwf,fayet,nl} 
on low-energy goldstino interactions. Focussing on the terms 
relevant for our calculation, we obtain 
\be
{{\cal L}}_{eff} = 
{1 \over F^2} [ \partial_{\mu} ( \psie \psis )]
[ \partial^{\mu} ( \psieb \psisb ) ] + \ldots \, .
\label{leff}
\ee

Could we have derived (\ref{leff}) from the known results on 
non-linear realizations? To address this question, we recall 
that the standard literature on the subject \cite{va,nl} 
prescribes an effective interaction of the form
\be
\label{leffbis}
{{\cal L}}_{eff}' = 
{i \over 2 F^2} [ \psis \sigma^{\mu} \partial^{\nu} \psisb
- ( \partial^{\nu} \psis ) \sigma^{\mu} \psisb ] \; T_{\nu \mu} 
+ \ldots \, ,
\ee
where $T_{\nu \mu}$ is the canonical energy-momentum tensor
of the matter fermions,
\be
T_{\nu \mu} = i \psieb \ov{\sigma}_{\nu} \partial_{\mu}
\psie + \ldots \, .
\label{timunu}
\ee
Combining (\ref{leffbis}) with (\ref{timunu}), we obtain
a result that looks very different from (\ref{leff}): 
\be
\label{leffter}
{{\cal L}}_{eff}' = 
- {1 \over F^2}  (\psis \sigma^{\mu} \partial^{\nu} \psisb) 
(\psieb \ov{\sigma}_{\nu} \partial_{\mu} \psie) + \ldots \, .
\ee

To check that (\ref{leff}) and (\ref{leffter}) are really 
inequivalent, we can concentrate on the process $f \; \ov{f} 
\to \grav \grav$. The only non-vanishing amplitudes are 
those in which the  two goldstinos have opposite 
chiralities. From eq.~(\ref{leff}), and in obvious 
notation, we obtain:
\be
\label{amp}
a(L,R,L,R) = -{(1 +\cos \theta)^2 s^2 \over 4 F^2} \, ,
\;\;\;\;
a(L,R,R,L) =  {(1 -\cos \theta)^2 s^2 \over 4 F^2} \, .
\ee  From eq.~(\ref{leffter}) we obtain instead:
\be
\label{ampter}
a'(L,R,L,R) = {\sin^2 \theta s^2 \over 4 F^2} \, ,
\;\;\;\;
a'(L,R,R,L) = - {\sin^2 \theta s^2 \over 4 F^2} \, .
\ee
We conclude that the effective interactions (\ref{leff}) 
and (\ref{leffter}) give the same energy dependence, but  
different angular dependences (and total cross-sections).
Surprisingly, the two approaches lead to different results.

\section{Solution of the puzzle}

To understand the origin of the discrepancy, we apply to 
the present case the superfield construction of the non-linear 
realization of \cite{va,nl}. We define the superfield
\be
\label{bigl}
\Lambda_{\alpha} ( x,\theta,\ov{\theta} ) 
\equiv \exp (\theta Q + \ov{\theta} \ov{Q} ) \, 
\grav_{\alpha} (x) 
=  \grav_{\alpha} + \sqrt{2} F \theta_{\alpha} +
{i \over \sqrt{2} F} ( \grav \sigma^{\mu} \ov{\theta} 
- \theta \sigma^{\mu} \ov{\grav} ) \partial_{\mu} 
\grav_{\alpha} + \ldots  \, ,
\ee
whose lowest component is the goldstino $\grav$, and the superfield
\be
\label{bige}
E_{\alpha} ( x,\theta,\ov{\theta} )  \equiv 
\exp (\theta Q + \ov{\theta} \ov{Q} ) \, f_{\alpha} (x)
=  f_{\alpha} + {i \over \sqrt{2} F} ( \grav \sigma^{\mu} 
\ov{\theta} - \theta \sigma^{\mu} \ov{\grav} ) \partial_{\mu} 
f_{\alpha} + \ldots  \, ,
\ee
whose lowest component is the matter fermion $f$. In the 
simple case under consideration, the goldstino couplings to
matter in the non-linear realizations of \cite{va,nl} are
described by the supersymmetric lagrangian
\be
\label{invone}
{1 \over 4 F^4} \int d^4 \theta \; \Lambda^2 \ov{\Lambda}^2
\; i \ov{E} \ov{\sigma}^{\mu} \partial_{\mu} E \, , 
\ee  
which leads precisely to the result of eq.~(\ref{leffter}),
as can be explicitly checked.

The crucial question is now the following: are there other
independent invariants, besides (\ref{invone}), that can
contribute to our effective interaction? The answer is 
positive \cite{bfzeegg}, since we can also write:
\be
\label{invtwo}
{\alpha \over F^2} \int d^4 \theta \; \Lambda E \; 
\ov{\Lambda} \ov{E} \, , 
\ee
where $\alpha$ is an arbitrary dimensionless coefficient. 
The new invariant (\ref{invtwo}) gives, among other things, 
\be
\delta {\cal L}_{eff}' = 
{\alpha \over 4 F^2} (\psis \sigma^{\mu} \partial^{\nu} 
\psieb) (\psisb \ov{\sigma}_{\nu} \partial_{\mu} \psie)
+ \ldots \, .
\label{deltal}
\ee From the contact interaction of eq.~(\ref{deltal}),
we obtain the following amplitudes:
\be
\delta a'(L,R,L,R) = \alpha {(1 +\cos \theta) s^2 \over 8 F^2} \, ,
\;\;\;\;
\delta a'(L,R,R,L) = -\alpha {(1 -\cos \theta) s^2 \over 8 F^2} \, .
\label{ampl}
\ee
We may now wonder whether an appropriate linear combination of the
two invariants can reproduce the result of eq.~(\ref{amp}).
Indeed, it is immediate to check that, with the special choice 
$\alpha=-4$, the combination ${\cal L}_{eff}' + \delta {\cal L}_{eff}'$ 
reproduces the scattering amplitudes obtained from ${\cal L}_{eff}$:
the puzzle is solved!

As a first comment, we stress that there is no reason to believe 
that the result of eq.~(\ref{leff}) is more fundamental than the 
standard result of eq.~(\ref{leffter}). Since two independent 
invariants can be constructed, there is just an ambiguity in the 
effective theory description, parametrized by the coefficient 
$\alpha$ in eq.~(\ref{deltal}), which can be fixed only by the
underlying fundamental theory. At the level of the linear 
realization, this ambiguity is contained in the coefficients of 
higher-derivative operators, which are not included in the standard 
K\"ahler formulation of eq.~(\ref{expan}).

In summary, the low-energy theorems of supersymmetry, as expressed
in the standard literature on non-linear realizations \cite{va,nl}, 
must be modified: in the low-energy effective lagrangian describing 
the goldstino couplings to matter there are terms that, contrary to 
previous expectations, are not entirely controlled by the supersymmetry
breaking scale. At a second thought, this result is not as surprising 
as it may seem: there is a suggestive analogy with the textbook 
case of pion-nucleon scattering \cite{sw}, where the effective 
lagrangian consists of two independent terms, one fully controlled 
by the broken $SU(2) \times SU(2)$ symmetry, the other one 
containing the axial coupling $g_A$ as a free parameter.

Are (\ref{invone}) and (\ref{invtwo}) the only independent 
invariants that contribute to the effective four-fermion coupling 
under consideration, or are there others? After a straightforward
but tedious classification, it was found in \cite{bfzeegg} that,
assuming matter conservation, the parametrization of the most 
general on-shell amplitudes involving two goldstinos $\grav$ and 
two massless matter fermions $f$ does not require any additional 
invariant. 

It is of course interesting to generalize the above framework 
by including gauge interactions. At the level of local 
four-fermion operators, the previous result is not affected. 
In particular \cite{bfzeegg}, there are no $d=6$ local 
supersymmetric operators contributing to $e^+ e^- \to 
\grav \grav$ in the limit of vanishing electron mass and 
negligible selectron mixing. If present, these operators 
would have been characterized by a dimensionful coupling 
$M^2/F^2$, where $M$ is an independent mass scale, possibly 
arising from the underlying fundamental theory, and the 
process $e^+ e^- \to \grav \grav$ could have been used to 
extract interesting lower bounds on $m_{3/2}$ from 
supernova cooling, but this is not the case.

\section{Phenomenological implications}

When extended to observable processes and realistic models, the
previous results have other important phenomenological implications.
As discussed in \cite{bfz3}, powerful processes to search for 
a light gravitino $\grav$ (when the supersymmetric partners 
of the Standard Model particles and of the goldstino are above 
threshold) are $e^+ e^- \to \grav \grav \gamma$ and $q \ov{q} 
\to \grav \grav \gamma$, which would give rise to a distinctive 
$(photon + missing \; energy)$ signal. The first process can be 
studied at $e^+ e^-$ colliders such as LEP or the proposed NLC, 
the second one at hadron colliders such as the Tevatron or the 
LHC. At hadron colliders, we can also consider the partonic
subprocesses $q \ov{q} \to \grav \grav g$, $q g \to q \grav 
\grav$, $\ov{q} g \to \ov{q} \grav \grav$ and $g g \to g \grav 
\grav$, all contributing to the $(jet + missing \; energy )$ signal.

Consider, for definiteness, the reaction $e^+ e^-  \to \grav \grav 
\gamma$. As before, we can start from a linear realization with 
pure $F$-breaking, neglecting higher-derivative terms and selectron 
mixing. Also in this case, explicit integration of the heavy 
superpartners gives results \cite{bfz4} that differ from 
those obtained \cite{nach} in the non-linear realization 
of \cite{va,nl}. For a model-independent study, we 
would need the general form of the low-energy effective 
interactions, allowed by the non-linearly realized supersymmetry, 
that may contribute to the relevant amplitudes at leading order. 
This was not known at the time of the Workshop, but a definite
theoretical prescription for such an investigation has become 
available in the meantime \cite{clnew} (the same paper has also 
confirmed the results of \cite{bfzeegg}, and found the explicit 
general form of the coupling of two on-shell goldstinos to a single 
photon). In the absence of a general phenomenological analysis, 
we can use the results of \cite{bfz4} and derive from the present 
LEP data the tentative lower bound $m_{3/2} > 10^{-5} \ev$.
Because of the strong and universal power-law behaviour of the 
cross-section, always proportional to $s^3/|F|^4$, and the
absence of accidental zeroes, this bound is expected to be rather 
stable with respect to variations of the parameters characterizing
the most general non-linear realization. However, should a signal 
show up at LEP or at future linear colliders, having the general 
expression of the cross-section would be very important, since a 
detailed analysis of the photon spectrum would offer the unique 
opportunity of distinguishing among possible fundamental theories.

At hadron colliders, the analysis is more complicated, since
different signals can be considered, with several partonic
subprocesses contributing, and the background is more severe. 
However, the prospects for the present Tevatron data are at 
least as good as for LEP, since the higher available energy
is more than enough to compensate for the more difficult
experimental environment \cite{bfmz}. The present Tevatron
bound is estimated to be $m_{3/2} > 2.7 \times 10^{-5} \ev$,
and the LHC should be sensitive to values of $m_{3/2}$ up
to $1.2 \times 10^{-3} \ev$.

In summary, high-energy colliders are by far the best environment
to test the possible existence of a very light gravitino. In our 
opinion, it would be important to provide our experimental 
colleagues with a model-independent framework for such a search,
and we hope to complete this project soon.

\section*{Acknowledgements}

The speaker would like to thank A.~Brignole and F.~Feruglio
for the pleasant and stimulating collaboration on which the
original part of this talk is based. This work was supported
by the European Commission TMR Programme ERBFMRX-CT96-0045. 

\end{document}